\documentclass{article}

\usepackage{microtype}
\usepackage{graphicx}
\usepackage{dblfloatfix}  
\usepackage{subfigure}
\usepackage{float}
\usepackage{enumitem}
\usepackage{stackengine}
\usepackage{booktabs} 

\usepackage{hyperref}

\usepackage[accepted]{icml2021}

\icmltitlerunning{Multi-Genre Music Transformer - Composing Full Length Musical Piece}

\begin{document}

\twocolumn[
\icmltitle{Multi-Genre Music Transformer - Composing Full Length Musical Piece}



\icmlsetsymbol{equal}{*}

\begin{icmlauthorlist}
\icmlauthor{Abhinav Kaushal Keshari (Purdue University)}{}
\end{icmlauthorlist}

\icmlkeywords{Artificial Intelligence}

\vskip 0.3in
]

\begin{abstract}
In the task of generating music, the art factor plays a
big role and is a great challenge for AI. Previous
work involving adversarial training \cite{dong2018musegan} to
produce new music pieces and modeling the compatibility
\cite{huang2021modeling} of variety in music (beats, tempo, musical
stems) demonstrated great examples of learning this
task. Though this was limited to generating mashups or learning features from tempo and key distributions to produce similar patterns. Compound Word Transformer \cite{hsiao2021compound} was able to represent music generation task as
a sequence generation challenge involving musical events
defined by compound words. These musical events give a more accurate description of notes progression, chord change, harmony and the art factor. The objective of the
project is to implement a Multi-Genre Transformer which learns to produce
music pieces through more adaptive learning process
involving more challenging task where genres or form of the
composition is also considered. We built a multi-genre compound word dataset, implemented a linear transformer \cite{katharopoulos2020transformers}
which was trained on this dataset. We call this Multi-Genre Transformer, which was able to generate full length new musical pieces which is diverse and comparable to original tracks. The model trains 2-5 times faster than other models discussed.
\end{abstract}

\section{Related Work}

Despite achieving great success in generation challenges
using Artificial Intelligence in Natural Language Generation
(NLG) there is a factor of art that still makes them different
from human like performance. In terms of NLG we can
relate it to something like the difference between computer
generated article and a piece of art like novels, biography,
etc. For music art factor always come into account and despite able to produce musical compositions through Adversarial networks or mixing stems using supervised learning the solution still is very different from an original piece of music which we discuss below.

\subsection{Music Generation using GANs}

Generative adversarial networks (GANs) have provided significant progress in producing text, videos and
images. Similar efforts have been made to bring neural
networks to artistic domain of music. MuseGAN\cite{dong2018musegan} brought a
novel model for generating multi-track music.

Until 2018, the progress in using AI to compose music had
been able to produce

\begin{itemize}[noitemsep]
    \item Single-track (monophonic) music
    \item Multi-track (polyphonic) music by combining several monophonic melodies in chronological order
\end{itemize}

Music usually being an art involving multiple instruments
played together requires music to be multi-track and because
music notes are made up of chords, arpeggios or melodies
the idea of using a chronological order setting prevents it
from being generalized.

The paper\cite{dong2018musegan} address this challenge in generalising real music
by discussing current technical lacks in neural network
models and how it relates to the real world music.

\begin{enumerate}
\item Music is an art of time and has characteristics of coherence,
rhythm, tension and emotion flow. This requires
it to have a \textbf{Temporal Model}.
\item Music compositions usually involves different instruments
interacting with one another making the compositions
to be harmonic. To solve this issue a \textbf{Composer
Model} is required.
\item Musical notes are built of chords, arpeggios or
melodies and how they unfold over time; thus introducing
chronological generation of notes is not suitable.
To address this the paper introduces using bars (segment
of time) instead of notes as the basic unit for
composition. And then generate music bar by bar using
transposed convolutional neural networks to learn
translation-invariant patterns.
\end{enumerate}

The paper\cite{dong2018musegan} makes contributions in terms of both ability to
artificially compose realistic music and use of generative adversarial
framework with temporal and composition models.
In short the contributions are:

\begin{itemize}[noitemsep]
    \item First GAN based model for generating multi-track sequence.
    \item First model which can generate multi-track polyphonic
music.
    \item Same model can be used as a music accompaniment.
    \item Creates a new Lakh Pianoroll Dataset (LPD) for multitrack
piano-rolls
    \item For future work metrics in the domain of artificial music
a new set of objective metrics are proposed.
\end{itemize}

MuseGAN model proposed considers two sub-network generator Gtemp (temporal structure
generator) and Gbar (bar generator) making the overall
generator:
\begin{center}
    $G(z) = \left \{ G_{bar}(G_{temp}(z)^{(t)})  \right \}^{T}_{t=1}$
\end{center}
 where z is the
input noise vector. The strength of the model is the ability
to generate samples having chord like intervals (learning
features from temporal model) and melodies involving pitch
overlap among guitar, piano and strings (learning features
from composer model).

The model introduces
multi-track by modeling interdependency of tracks by
proposing 3 different generator model (Jamming, Composer
and Hybrid), but the author brings up these based on the
understanding of pop music composition. This possibly restricts the generator
to explore on a broad spectrum of music and prevents it from being generalised. Also worth mentioning is that the work relies on multi-track interdependency, but misses to
study about the compatibility of these tracks which can significantly increase the quality of music being generated. We will see this issue being addressed in the next paper.

\subsection{Modeling the Compatibility of Stem Tracks
to Generate Music Mashups\cite{huang2021modeling}}

Source separation\cite{jansson2017singing, defossez2019music} makes it possible to generate
a music mashup with isolated stems like vocals, drums,
piano, etc. The challenge lies in producing music which
has compatibility between these stems. This paper creates a
mashup generation pipeline and trains a model to predict the
compatibility by automatically learning to adjust key and
tempo (characteristics of quality mashups in real world).

General models trained for harmonic compatibility \cite{bernardes2017hierarchical, macas2018mixmash} fails to consider subtle features or surprise mixes of disparate samples which is quite common in this art domain.
Other issue that arises is audio compatibility models like
Neural Loop Combiner \cite{chen2020neural} having lack of
vocal source and variety of genres.

The authors designed a \textbf{self supervised learning} model by recombining
the original combination of stems before source separation
to serve as examples of ground truth. To avoid highly polarized
model, \textbf{semi-supervised learning} was introduced
which included producing several random mashups by mixing
different stems and treated them as unlabeled instances.
Label smoothing regularization for outliers \cite{zheng2017unlabeled} was used to assign uniform distribution to the unlabeled
data for loss computation. This helps in regularization.

The final architecture consists of 3 modules:

\begin{enumerate}
    \item \textbf{Music Source Separation:} Uses MSS algorithm \cite{jansson2017singing} to get different stems \textit{vocals},
\textit{drums}, \textit{bass} and \textit{other}.
    \item \textbf{Mashup Database (MashupDB):} Using Madmom \cite{bock2016madmom} different features from the music
clips are extracted like \textit{key}, \textit{tempo} and \textit{downbeat} information.
Using these features and separate stem
combinations a mashup database is created which will
act as either harmonic or percussion stem candidates
for mashup generation process.
    \item \textbf{Mashup Generation:} It uses candidate stems from
MashupDB and adjusts key and tempo to produce
mashups within 3 conditions - original, matched and
unmatched.
\end{enumerate}

The model \cite{huang2021modeling} is defined by $p(y|V,H,P)$ where $V$, $H$, and $P$
are input signals for respective stems \textit{vocal}, \textit{harmonic}, and
\textit{percussion}. The output probability p is used as the mashup
compatibility and $y \in \left \{ 0, 1 \right \} $ stating good or bad.

The  model \cite{huang2021modeling} implementation tries to mimic learning compatibility
for producing new mashups and provides objective
and subjective evaluation by cross validation among multiple
different datasets. This technique becomes easier because
of the ability of the model to extract different stems
and features and build its own mashup candidates. This also
makes the model training process not dependent on human
labeled data. The model is also robust as negative data is
added along with positive data for supervised learning. The
range of music coverage is also extensive and the source
separation step makes it easier for the model to be extended
to different genres for training.

But the current model design lacks the effective embedding
of different stems while producing a mashup and makes
it dependent on tuning of key and tempo. Currently the
implementation comes up with fixed range of key and tempo
difference for compatibility and does not explain in detail
how they came up with these numbers. Although defining
a range prevents large pitch shifting and time stretching.
Additionally the results of the model ranks positive labeled data (original) over unlabeled data which might lead to
concerns of flexibility. Another major challenge of the
model is the large training time which is around 3 days
using an NVIDIA Tesla-V100 GPU whereas using transformer model significantly reduces the training time.

\subsection{Music Transformers}

With state-of-the art neural network we managed to learn
features in music by defining certain rules on matching
tempo, beats or compatibility. In the previous paper we
also tried to learn compatibility with the help of supervised
learning. The model though suffered with bias as compatibility
was favoured for matched key or tempo and also lacks
generalization. Compound Word Transformer \cite{hsiao2021compound} considers music as sequence of
events and uses a Transformer (neural sequence model) \cite{vaswani2017attention} to
generate a new musical sequence.

A musical note can be described by note’s pitch, chord, bar,
duration, velocity (dynamics), placement (onset time). If
we consider these as tokens we can then define music as
sequence of tokens and these tokens are a part of pre-defined
vocabulary. As music is multi-faceted a particular type of
token can capture only a certain feature like melody, rhythm,
harmony. All the neural networks until now treated these
tokens as equal and thus lacked heterogeneity.

Compound Word Transformer \cite{hsiao2021compound} generates music in a conceptually different way
as it allows tokens to be of specific types and let them have
their own properties. Tokens can be of note type (pitch,
duration) or metric type (beginning of new beat, bar). We
then defines a musical event by combination of such tokens
which allows to capture co-occurrence relationship
among the tokens. This combination of tokens are termed
as compound words. So, now we can represent a music
piece $(X)$ as a sequence $(S)$ of compound words $(cp)$ or
$S = g(X) = \left \{ cp_t\right \}^{T}_{t=1}$ where $g(.)$ is the conversion function
to convert music into time-ordered sequence of musical
events and $T$ is the length of the music sequence.

\begin{figure}[ht]
\vskip 0.2in
\begin{center}
\centerline{\includegraphics[width=\columnwidth]{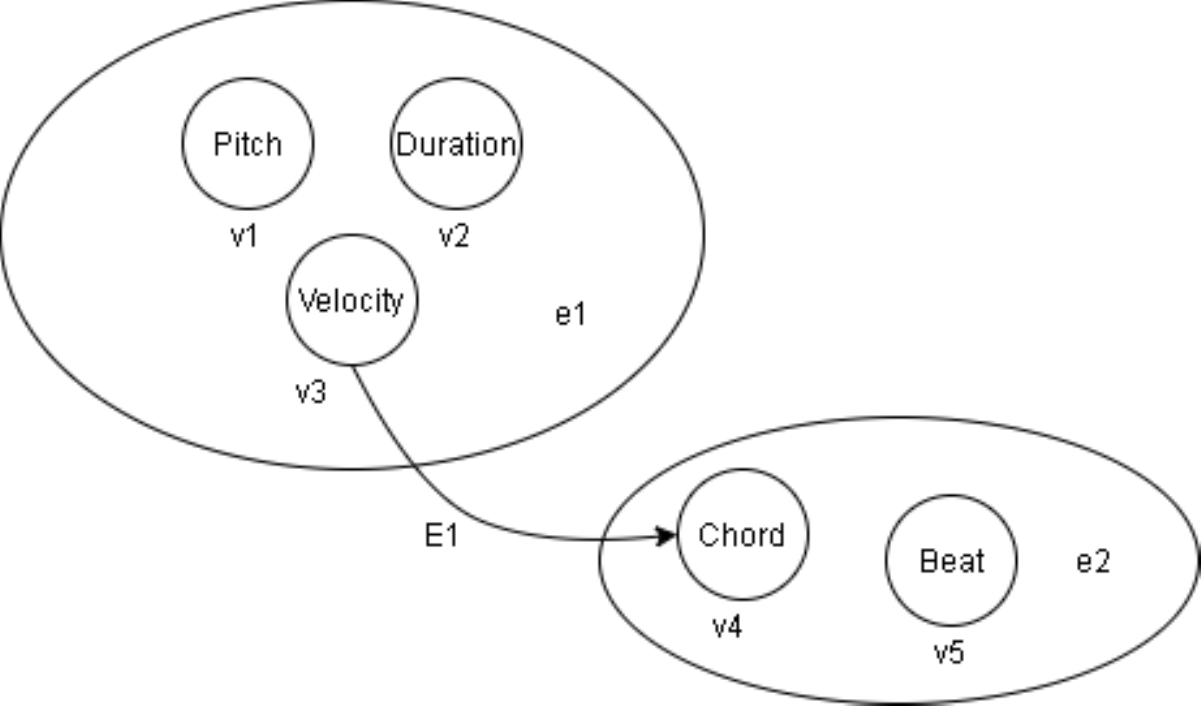}}
\caption{Graphical Representation of Music Space}
\label{icml-historical}
\end{center}
\vskip -0.2in
\end{figure}

Theoretically, the model learns over discrete-time dynamic
directed hypergraphs. Consider a graph $G = (V,E)$ (Figure
1) the vertices $(V)$ are tokens and edges $(E)$ are sequence of
token. Collection of vertices can be defined as a compound
word and hyperedge in this graph represents sequence of
compound words. In figure 1 $v1, v2, v5$ are the tokens and
the edge $E1$ defines a sequence of tokens whereas $e1, e2$
defines a hyperedge (connecting more than 2 nodes). And
transitioning from one hyperedge to another defines the
sequence of composition words which we are trying to learn.

Using a transformer we are trying to learn the next musical event or compound word (combination of tokens). The self attention part of the transformer learns the dependency
among the elements in musical sequence and different feedforward
head is used for tokens of different type. In short
the implementation groups tokens to form compound words
and then perform sequence modeling in this sequence of
compound words, the major contributions are:

\begin{itemize}[noitemsep]
    \item Compose pop-piano music of full song length.
    \item Compound word sequencing with linear transformer
providing state-of-the-art results in terms of quality
with 5-10x faster training and inference time.
    \item Music defined as Dynamic Directed Hypergraph.
\end{itemize}

Generating a new musical event or a group of tokens to
be combined as a compound word at each time step is the
backbone of this model, but it relies on assuming that no
two musical events can occur together. The new hyperedge
generated by the Transformer decoder marks other tokens as
[ignore] once an event of a particular token type is detected.
Can this limit the music generation task? Additionally the model is trained using only pop music which limits the expressing power of the transformer.

\section{Implementation}

Compound Word Transformer \cite{hsiao2021compound} was able to represent music generation task as
a sequence generation challenge involving musical events
defined by compound words. Leveraging this representation we implement a neural model which learns to produce
music pieces through more adaptive learning process
involving more challenging task where genres or form of the
composition is also considered. This adds the richness of
music art in the learning process of attention driven sequential
learning. We will call this model \textit{Multi-Genre Music Transformer} and following are the steps involved for implementing this:

\begin{itemize}[noitemsep]
    \item \textbf{Building Dataset}: This involves generating compound word dictionary for songs of different genres.
    \item \textbf{Implementing Transformer Model}: We implement our Transformer class, the training steps and the generation logic for inference.
    \item \textbf{Adaptive Learning}: We allow our tuned model to be adaptable by training on a smaller and multi-genre dataset.
\end{itemize}

\subsection{Building Dataset}
To be able to provide a more generalised learning process for our transformer it needs to be trained with a piano roll dataset involving musical pieces of variety of genres/style. The dataset should be based on compound words \cite{hsiao2021compound} to represent different musical tokens as a combined unit for sequence modeling which is different from traditional musical dataset (MIDI, REMI).

\begin{figure}[hbt!]
\begin{center}
\centerline{\includegraphics[width=\columnwidth]{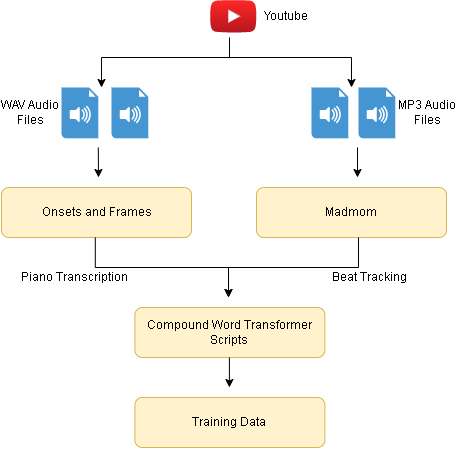}}
\caption{Dataset Building Pipeline}
\end{center}
\vskip -0.2in
\end{figure}

This required us to build a dataset by selecting music clippings
and converting them to piano roll using Onsets and
Frames \cite{hawthorne2017onsets}. Extracting downbeat and
beat information from these songs using madmom, a music
signal processing library \cite{bock2016madmom}. Finally
representing these metadata into a compound word representation
using the dataset generation scripts provided in the
compound word transformer repository\footnote{https://github.com/YatingMusic/compound-word-transformer/blob/main/dataset/Dataset.md}. This also adds on to the AILabs.tw Pop1K7 dataset \cite{hsiao2021compound} which currently only includes pop music. Figure 2 demonstrates the pipeline for creating a new dataset.

Following the pipeline above we managed to create a Compound Word \cite{hsiao2021compound} dataset which involved piano roll for 150 musical pieces from 3 different genres including Electronic Dance Music (EDM), Indie and Hip-Hop.

\subsection{Implementing Transformer Model}
We implement a linear transformer\cite{katharopoulos2020transformers} to address long
sequence dependency which is a very relevant factor in
music generation due to the presence of a context or a rhythm
in the entire musical piece. Having an independent feed-forward head in the Transformer Decoder allows to improve the loss of independent tokens. This allows
the model to scale for additional perspective (like genre, form or involving a particular
chord progression) in the music by adding an additional token type. We implement our transformer model in a generic way which allows user to define its own token sampling model, token embedding model and these
can be scalable for any number of token types. The loss observed at each
feed-forward head is shown in Figure 6. This shows adding a new token (for genre/style/form) for model to learn can be simply achieved by adding an independent feed-forward head for the same.

\subsubsection{Token Embedding}

\begin{figure}[H]
\begin{center}
\centerline{\includegraphics[width=\columnwidth]{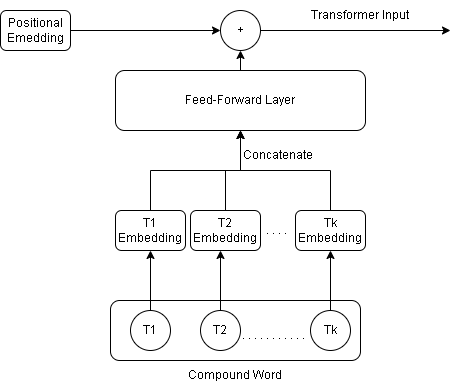}}
\caption{Demonstrates how each token undergoes independent embedding before combining with Positional Encoding. Here $T_1, T_2 ... T_k$ are $K$ different tokens for our Transformer each having its own embedding function and dimension. We are assuming the Transformer supports $K$ type of tokens.}
\end{center}
\vskip -0.2in
\end{figure}

The input to a transformer requires positional encoding added to the embedding vector of our input sequence elements. As each element in our sequence is a compound word \cite{hsiao2021compound} which is combined of different tokens, we embed each token separately (allowing to have adaptive size) and then concatenate them. Having an adaptive token size allows to use smaller embedding dimension for a token type with smaller vocabulary and when we concatenate all of these we get an embedding dimension of 512 for our model. Refer to Figure 3 for detailed steps of token embedding.

\subsubsection{Token Sampling}
For inference, sampling plays a crucial role to avoid degeneration and improve diversity. To avoid degeneration we follow Nucleus Sampling \cite{holtzman2019curious}, which is a stochastic temperature controlled process. This method samples from the smallest subset of tokens whose cumulative probability mass exceeds a threshold. We also had each token to have a separate sampling policy by defining different threshold $p$ and different temperature parameter $\tau$ \cite{ackley1985learning} for reshaping the probability before sampling. We reused the inference implementation from Compound Word Transformer \cite{hsiao2021compound} and tweaked $\tau$ to have higher values for chord to allow more diverse chord progressions. 
Figure 4 shows the sampling process and individual feed-forward layer for each token in the transformer.

\begin{figure}[H]
\begin{center}
\centerline{\includegraphics[width=\columnwidth, height=10cm]{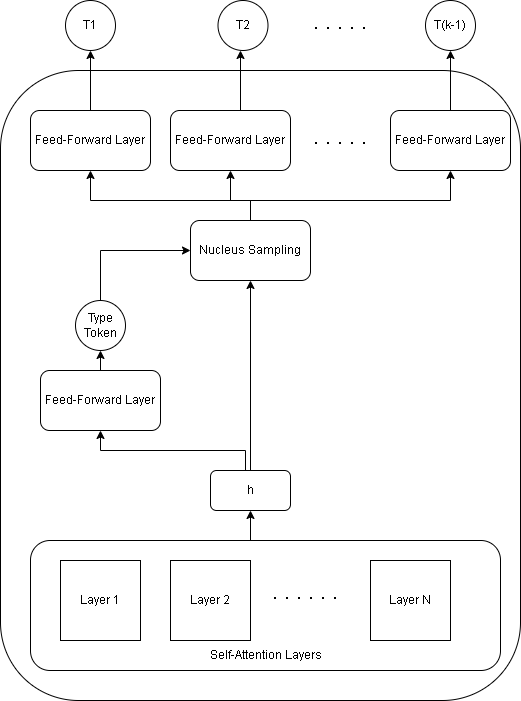}}
\caption{Transformer with N self-attention layers and independent feed-forward head for each token. We first predict the Token Type for the particular time-step and then perform a nucleus sampling before predicting the remaining tokens.}
\end{center}
\vskip -0.2in
\end{figure}

\subsection{Adaptive Learning}
After defining the model, the next important step is to implement the training steps. To support scalable token definition in our generalised transformer we make the training steps modular and general to variable number of token types. This allows easy addition of a new token and independently monitor gradient descent optimization for the respective loss.

We trained our model in parallel for 2 different conditions. The first set of training was performed on the original AILabs.tw Pop1K7 dataset \cite{hsiao2021compound}. The second set of training took into consideration to provide multi-genre learning environment for the transformer as it involved training on a dictionary that was generated from 3 different genres (EDM, Indie, Hip-Hop).

\section{Evaluation and Results}
To train a multi-genre transformer the primary objective was to provide it with a dataset which is richer in variety than the original pop only dataset. With the help of dataset building pipeline we managed to create a token set which has a higher variance allowing the model to have a broader expressive power. Figure 5 shows the comparison of tokens between the 2 datasets used.

\begin{figure}[H]
\begin{center}
\centerline{\includegraphics[width=\columnwidth, height=4cm]{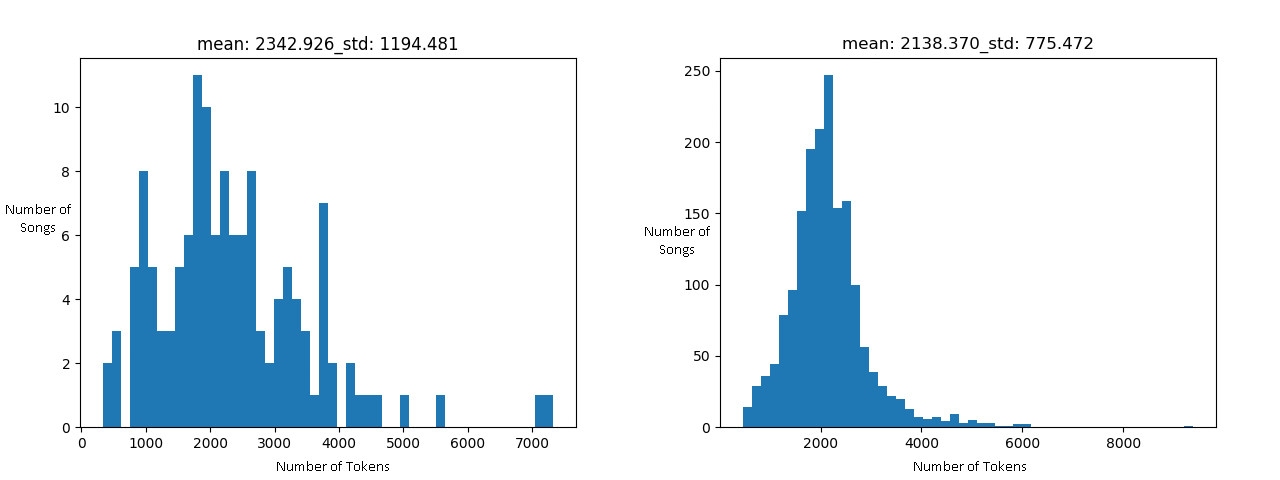}}
\caption{Left image shows token distributions for the songs in the generated multi-genre dataset and the right image shows similar distribution for AILabs.tw Pop1K7 dataset \cite{hsiao2021compound}.}
\end{center}
\vskip -0.2in
\end{figure}

After training the model for both the datasets we also observe (refer to Figure 6) the individual token loss and total average loss is similar and indicates the model converging. Additionally, the gradient descent is more gradual using the multi-genre dataset displaying a more settled progression.

We trained the model with 12 self-attentions layers, 8 feed-forward heads with model dimension of 512 and batch size of 4 for 180 epochs which took around 17hrs. Then using the trained model we generated 20 new full length musical pieces with an average inference time of 12.56sec/song which is faster than the compound-word transformer though having slightly less number of average tokens per song. Table 1 shows a more detailed comparison.

\begin{figure*}[!t]
    \centering
    \includegraphics[width=\textwidth,height=8cm]{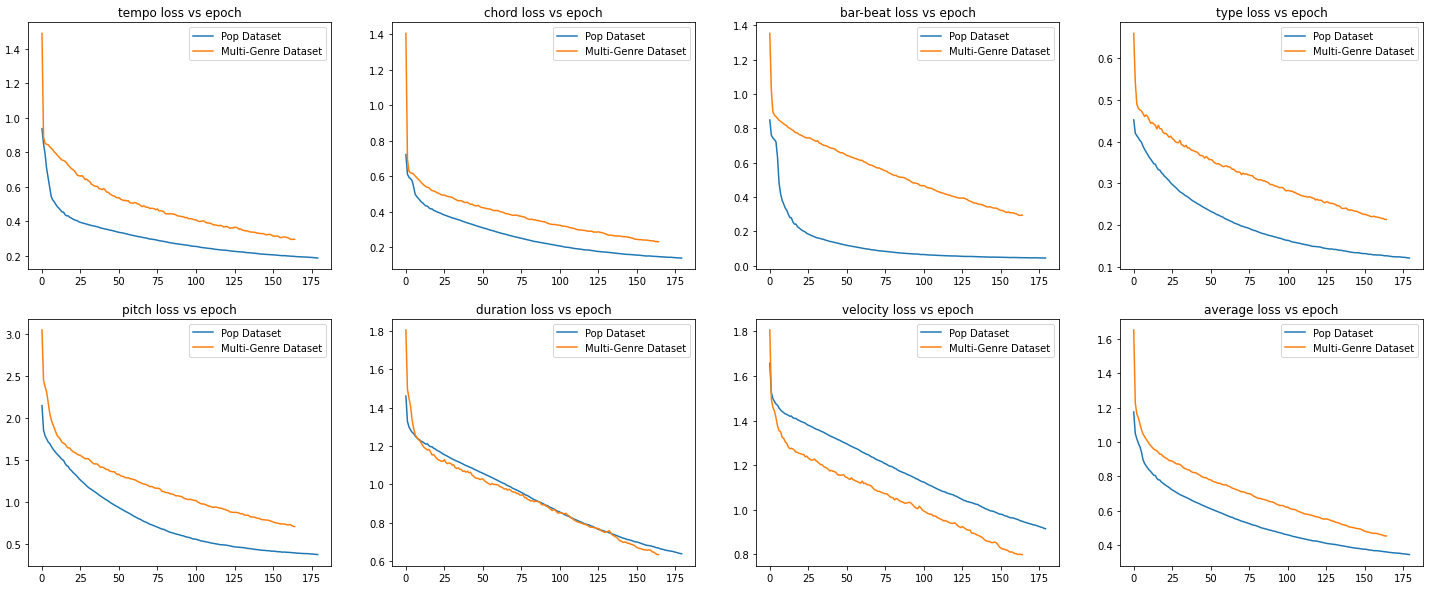}
    \caption{Loss vs Epoch for different token types. The last plot corresponds to the average loss for all different token types.}
\end{figure*}

\begin{table*}[b]
\caption{Quantitative evaluation results for Multi-Genre Transformer and Compound Word Transformer. Results for Compound Word Transformer comes from the implementation in the paper \cite{hsiao2021compound}.}
\label{sample-table}
\vskip 0.15in
\begin{center}
\begin{small}
\begin{sc}
\begin{tabular}{cccccc}
\toprule
Model & Training Time & GPU  & Inference Time (/song) & Avg Tokens (/song) \\
\midrule
Multi-Genre Transformer    & 17 hrs & 9.8GB  & 12.56 sec & 9190 \\
Compound Transformer    & 1.3 days & 9.5GB  & 19.8 sec & 9546 \\
\bottomrule
\end{tabular}
\end{sc}
\end{small}
\end{center}
\vskip -0.1in
\end{table*}

For a qualitative evaluation of the musical pieces that were produced we compare (Figure 7) the piano rolls of these with the piano rolls of original tracks that were used to train the model.

\begin{figure}[H]
\vskip 0.2in
\begin{center}
\stackunder[3pt]{\includegraphics[width=\columnwidth,height=1in]{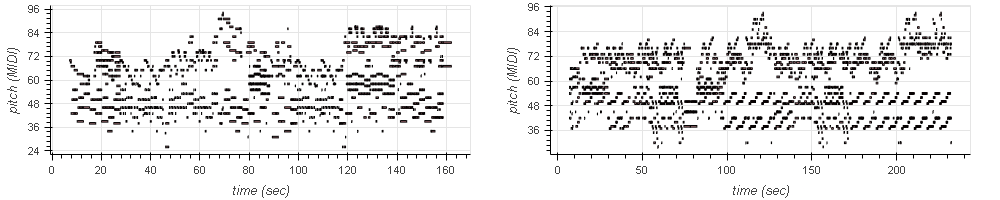}}{Original Songs}%
\newline
\stackunder[3pt]{\includegraphics[width=\columnwidth,height=1in]{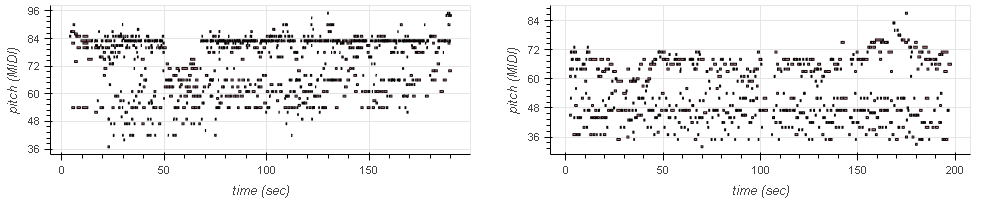}}{Generated Songs}%
\caption{Piano roll of original and generated songs. We can see a rich and complete content for the generated songs similar to some original tracks.}
\end{center}
\vskip -0.2in
\end{figure}

\section{Conclusion}
In this project we produce music as a sequence of musical events produced by a trained Transformer. We leverage the definition of Compound Word \cite{hsiao2021compound} to define musical event by grouping multiple tokens. This grouping greatly reduces the size of our sequence and boosts long-range learning. This also reduces the training and inference time for our model remarkably. We also exploit the feature of each token having its independent feed-forward head for prediction to make the model scalable for new token types that can be introduced in our dictionary. This allows to add any new token for this transformer very easily which can be used for musical form, chord progression, etc. Additionally, we created an entire new dataset consisting of multi-genre compound word dictionary and trained our model with this to provide it a more adaptive learning environment. The compositions that were generated were highly rich in musical events and were of good quality.


\bibliography{example_paper}

\begin{thebibliography}{15}
\providecommand{\natexlab}[1]{#1}
\providecommand{\url}[1]{\texttt{#1}}
\expandafter\ifx\csname urlstyle\endcsname\relax
  \providecommand{\doi}[1]{doi: #1}\else
  \providecommand{\doi}{doi: \begingroup \urlstyle{rm}\Url}\fi

\bibitem[Ackley et~al.(1985)Ackley, Hinton, and Sejnowski]{ackley1985learning}
Ackley, D.~H., Hinton, G.~E., and Sejnowski, T.~J.
\newblock A learning algorithm for boltzmann machines.
\newblock \emph{Cognitive science}, 9\penalty0 (1):\penalty0 147--169, 1985.
\newblock URL
  \url{https://www.sciencedirect.com/science/article/abs/pii/S0364021385800124}.

\bibitem[Bernardes et~al.(2017)Bernardes, Davies, and
  Guedes]{bernardes2017hierarchical}
Bernardes, G., Davies, M.~E., and Guedes, C.
\newblock A hierarchical harmonic mixing method.
\newblock In \emph{International Symposium on Computer Music Multidisciplinary
  Research}, pp.\  151--170. Springer, 2017.
\newblock URL
  \url{https://link.springer.com/chapter/10.1007/978-3-030-01692-0_11}.

\bibitem[B{\"o}ck et~al.(2016)B{\"o}ck, Korzeniowski, Schl{\"u}ter, Krebs, and
  Widmer]{bock2016madmom}
B{\"o}ck, S., Korzeniowski, F., Schl{\"u}ter, J., Krebs, F., and Widmer, G.
\newblock Madmom: A new python audio and music signal processing library.
\newblock In \emph{Proceedings of the 24th ACM international conference on
  Multimedia}, pp.\  1174--1178, 2016.
\newblock URL \url{https://dl.acm.org/doi/abs/10.1145/2964284.2973795}.

\bibitem[Chen et~al.(2020)Chen, Smith, and Yang]{chen2020neural}
Chen, B.-Y., Smith, J.~B., and Yang, Y.-H.
\newblock Neural loop combiner: Neural network models for assessing the
  compatibility of loops.
\newblock \emph{arXiv preprint arXiv:2008.02011}, 2020.
\newblock URL \url{https://arxiv.org/abs/2008.02011}.

\bibitem[D{\'e}fossez et~al.(2019)D{\'e}fossez, Usunier, Bottou, and
  Bach]{defossez2019music}
D{\'e}fossez, A., Usunier, N., Bottou, L., and Bach, F.
\newblock Music source separation in the waveform domain.
\newblock \emph{arXiv preprint arXiv:1911.13254}, 2019.
\newblock URL \url{https://arxiv.org/abs/1911.13254}.

\bibitem[Dong et~al.(2018)Dong, Hsiao, Yang, and Yang]{dong2018musegan}
Dong, H.-W., Hsiao, W.-Y., Yang, L.-C., and Yang, Y.-H.
\newblock Musegan: Multi-track sequential generative adversarial networks for
  symbolic music generation and accompaniment.
\newblock In \emph{Thirty-Second AAAI Conference on Artificial Intelligence},
  2018.
\newblock URL
  \url{https://www.aaai.org/ocs/index.php/AAAI/AAAI18/paper/viewPaper/17286}.

\bibitem[Hawthorne et~al.(2017)Hawthorne, Elsen, Song, Roberts, Simon, Raffel,
  Engel, Oore, and Eck]{hawthorne2017onsets}
Hawthorne, C., Elsen, E., Song, J., Roberts, A., Simon, I., Raffel, C., Engel,
  J., Oore, S., and Eck, D.
\newblock Onsets and frames: Dual-objective piano transcription.
\newblock \emph{arXiv preprint arXiv:1710.11153}, 2017.
\newblock URL \url{https://arxiv.org/abs/1710.11153}.

\bibitem[Holtzman et~al.(2019)Holtzman, Buys, Du, Forbes, and
  Choi]{holtzman2019curious}
Holtzman, A., Buys, J., Du, L., Forbes, M., and Choi, Y.
\newblock The curious case of neural text degeneration.
\newblock In \emph{International Conference on Learning Representations}, 2019.
\newblock URL \url{https://openreview.net/forum?id=rygGQyrFvH}.

\bibitem[Hsiao et~al.(2021)Hsiao, Liu, Yeh, and Yang]{hsiao2021compound}
Hsiao, W.-Y., Liu, J.-Y., Yeh, Y.-C., and Yang, Y.-H.
\newblock Compound word transformer: Learning to compose full-song music over
  dynamic directed hypergraphs.
\newblock In \emph{Proceedings of the AAAI Conference on Artificial
  Intelligence}, volume~35, pp.\  178--186, 2021.
\newblock URL \url{https://ojs.aaai.org/index.php/AAAI/article/view/16091}.

\bibitem[Huang et~al.(2021)Huang, Wang, Smith, Song, and
  Wang]{huang2021modeling}
Huang, J., Wang, J.-C., Smith, J.~B., Song, X., and Wang, Y.
\newblock Modeling the compatibility of stem tracks to generate music mashups.
\newblock In \emph{Proceedings of the AAAI Conference on Artificial
  Intelligence}, volume~35, pp.\  187--195, 2021.
\newblock URL \url{https://ojs.aaai.org/index.php/AAAI/article/view/16092}.

\bibitem[Jansson et~al.(2017)Jansson, Humphrey, Montecchio, Bittner, Kumar, and
  Weyde]{jansson2017singing}
Jansson, A., Humphrey, E., Montecchio, N., Bittner, R., Kumar, A., and Weyde,
  T.
\newblock Singing voice separation with deep u-net convolutional networks.
\newblock In \emph{18th International Society for Music Information Retrieval
  Conference}, pp.\  23--27, 2017.
\newblock URL
  \url{https://openaccess.city.ac.uk/id/eprint/19289/1/7bb8d1600fba70dd79408775cd0c37a4ff62.pdf}.

\bibitem[Katharopoulos et~al.(2020)Katharopoulos, Vyas, Pappas, and
  Fleuret]{katharopoulos2020transformers}
Katharopoulos, A., Vyas, A., Pappas, N., and Fleuret, F.
\newblock Transformers are rnns: Fast autoregressive transformers with linear
  attention.
\newblock In \emph{International Conference on Machine Learning}, pp.\
  5156--5165. PMLR, 2020.
\newblock URL \url{http://proceedings.mlr.press/v119/katharopoulos20a.html}.

\bibitem[Macas et~al.(2018)Macas, Rodrigues, Bernardes, and
  Machado]{macas2018mixmash}
Macas, C., Rodrigues, A., Bernardes, G., and Machado, P.
\newblock Mixmash: A visualisation system for musical mashup creation.
\newblock In \emph{2018 22nd International Conference Information Visualisation
  (IV)}, pp.\  471--477. IEEE Computer Society, 2018.
\newblock URL
  \url{https://www.computer.org/csdl/proceedings-article/iv/2018/720200a471/17D45XvMcd9}.

\bibitem[Vaswani et~al.(2017)Vaswani, Shazeer, Parmar, Uszkoreit, Jones, Gomez,
  Kaiser, and Polosukhin]{vaswani2017attention}
Vaswani, A., Shazeer, N., Parmar, N., Uszkoreit, J., Jones, L., Gomez, A.~N.,
  Kaiser, {\L}., and Polosukhin, I.
\newblock Attention is all you need.
\newblock In \emph{Advances in neural information processing systems}, pp.\
  5998--6008, 2017.
\newblock URL
  \url{https://proceedings.neurips.cc/paper/2017/file/3f5ee243547dee91fbd053c1c4a845aa-Paper.pdf}.

\bibitem[Zheng et~al.(2017)Zheng, Zheng, and Yang]{zheng2017unlabeled}
Zheng, Z., Zheng, L., and Yang, Y.
\newblock Unlabeled samples generated by gan improve the person
  re-identification baseline in vitro.
\newblock In \emph{Proceedings of the IEEE international conference on computer
  vision}, pp.\  3754--3762, 2017.
\newblock URL
  \url{https://openaccess.thecvf.com/content_iccv_2017/html/Zheng_Unlabeled_Samples_Generated_ICCV_2017_paper.html}.

\end{thebibliography}
\bibliographystyle{icml2021}


\end{document}